\documentclass[runningheads]{llncs}
\usepackage[T1]{fontenc}
\usepackage{color}
\usepackage{multirow}
\usepackage{bbding}
\usepackage{pifont}
\usepackage{adjustbox}
\usepackage{float}
\usepackage{graphicx}
\begin{document}
\title{
GCAN: Generative Counterfactual Attention-guided Network for Explainable Cognitive Decline Diagnostics based on fMRI Functional Connectivity
}
%
\titlerunning{Counterfactual Attention for fMRI Connectivity}
%
\author{Xiongri Shen\inst{1}
\and Zhenxi Song\inst{1} 
\and Zhiguo Zhang\inst{1} 
}
\authorrunning{X R. Shen et al.}
%
\institute{Department of Computer Science and Technology, Harbin Institute of Technology (Shenzhen), Shenzhen, China 
} %
%
%
\maketitle              
\vspace{-8mm}
\begin{abstract}
Diagnosis of mild cognitive impairment (MCI) and subjective cognitive decline (SCD) from fMRI functional connectivity (FC) has gained popularity, but most FC-based diagnostic models are black boxes lacking casual reasoning so they contribute little to the knowledge about FC-based neural biomarkers of cognitive decline. 
To enhance the explainability of diagnostic models, we propose a generative counterfactual attention-guided network (GCAN), which introduces counterfactual reasoning to recognize cognitive decline-related brain regions and then uses these regions as attention maps to boost the prediction performance of diagnostic models. Furthermore, to tackle the difficulty in the generation of highly-structured and brain-atlas-constrained FC, which is essential in counterfactual reasoning, an Atlas-Aware Bidirectional Transformer (AABT) method is developed. AABT employs a bidirectional strategy to encode and decode the tokens from each network of brain atlas, thereby enhancing the generation of high-quality target label FC.
In the experiments of hospital-collected and ADNI datasets, the generated attention maps closely resemble FC abnormalities in the literature on SCD and MCI. The diagnostic performance is also superior to baseline models. The code is available at \textcolor{blue}{https://github.com/SXR3015/GCAN}.
\vspace{-3mm}
\vspace{0mm}
\end{abstract}
\vspace{-9mm}
\section{Introduction}
\vspace{-3mm}
Diagnosing mild cognitive impairment (MCI) and subject cognitive decline (SCD) is vital for early intervention of Alzheimer's disease (AD). The fMRI-based functional connectivity (FC) abnormalities at network level \cite{Ramírez-Toraño2022} and/or region level \cite{Liebe2022FC} have been extensively in the research of MCI and SCD. 
FC is normally calculated as Pearson's Correlation Coefficients among fMRI signals on a set of predefined regions of interest (ROIs) derived from an anatomical or functional atlas. 
Based on FC, researchers have developed a great number of diagnostic models using deep learning methods, such as those based on convolution neural network (CNN) \cite{Li2021CNN} or Transformer \cite{Zuo2023t}. However, these models are normally a black-box so important FC and related brain regions that are predictive of SCD or MCI still remain unclear.
Some explainable models, such as Grad-CAM \cite{Selvaraju2016gradcam} and Score-CAM\cite{Wang2020scorecam}, have been popularly used in other fields like computer vision. However, they have generated explanation results using gradient backward based on the classification result labels. These methods often produce similar explanation results across incorrect class labels. Recently, counterfactual reasoning has emerged, creating the model's output in hypothetical scenarios, which directly generate explanation results, thus circumventing erroneous inference results based on result labels.

In the related research of MRI, Oh et al. applied counterfactual reasoning to structural MRI for the diagnosis of AD and MCI \cite{Oh2021cm}. Ren et al. also used counterfactual reasoning in the detection of brain lesions \cite{Ren2023cm}. 
However, applying counterfactual reasoning to FC presents challenges due to the strong structural characteristics of the brain atlas, making the reconstruction of the target FC difficult and thereby complicating the counterfactual reasoning architecture.


\begin{figure*}[ht]
\vspace{-5mm}
    \centering
    \includegraphics[scale=.5]{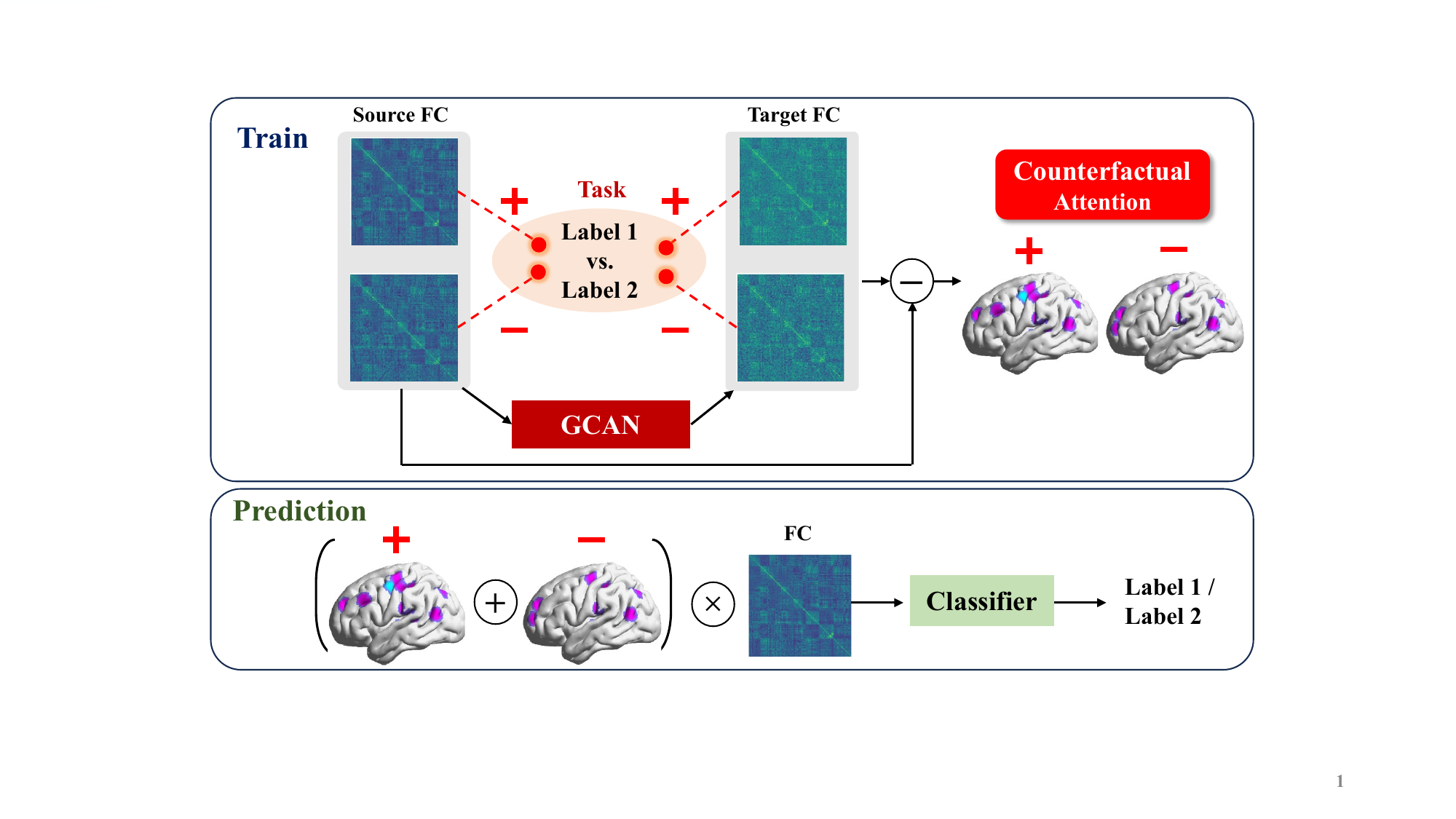}
    \vspace{-8mm}
    \caption{
    The proposed Generative Counterfactual Attention-guided Network (GCAN)
    interpretatively enhances the diagnosis of cognitive decline by identifying key brain regions (referred to as '\textit{counterfactual attention}') involved in the transition between healthy, SCD, and MCI brain states, based on their functional connectivity (FC) matrix.
    This approach offers a bidirectional view of counterfactual attention that, once integrated with the original FC matrix, guides the classifier's focus to these regions throughout the learning process. The positive (${+}$) and negative (${-}$) signs symbolize two perspectives within the attention map, reflecting the inversion of causal relationship positions-\textit{source} and \textit{target}-in counterfactual inference.}
    \label{fig: Fig1}
\end{figure*}

\vspace{-5mm}
To construct the counterfactual reasoning architecture for FC-based diagnostic models, we introduce the Generative Counterfactual Attention-Directed Network (GCAN) for identifying predictable FC features and related brain regions. These regions are then used as counterfactual attention maps onto FC matrices to increase the prediction performance. To tackle the challenge of generating FC, we devised an Atlas-aware Bidirectional Transformer (AABT) to reconstruct FC within the GCAN framework.
The main contributions of this paper are twofold:
\begin{itemize}
\vspace{-2mm}
   \item[1)] We introduce a counterfactual reasoning architecture to detect cognitive decline-related regions. 
   The architecture is shown in Fig. \ref{fig: Fig1}.   
    \begin{itemize}
        \item[$\bullet$]\textbf{training stage}: We develop GCAN to generate the target label FC. Subsequently, we compute the difference between the target label FC and the source label FC to construct the counterfactual attention for all source labels.
        \item[$\bullet$]\textbf{prediction stage}: The new diagnostic model is initialized with attention on cognitive decline-related regions by aggregating all counterfactual attention and applying the resulting total attention map to FC. Subsequently, this masked FC matrix is used to train the new diagnostic model.
   \end{itemize}
   \item[2)] AABT is introduced for generating highly structured FC. It can dynamically encode and decode the FC based on individual networks of atlas. To offer a better understanding of both encoding and decoding FC, we employ a bidirectional structure including forward and backward process to handle token encoding and decoding.
\end{itemize}


\vspace{-3mm}
\section{Method}

\subsection{Generative Counterfactual Attention-guided Network (GCAN)}

The proposed GCAN consists of a Generator and a Discriminator, each rooted in the principles of counterfactual inference. The fundamental architecture of both the generator and discriminator is the AABT, as detailed in Section 2.2.

\begin{figure*}[ht]
    \centering
    \includegraphics[scale=.42]{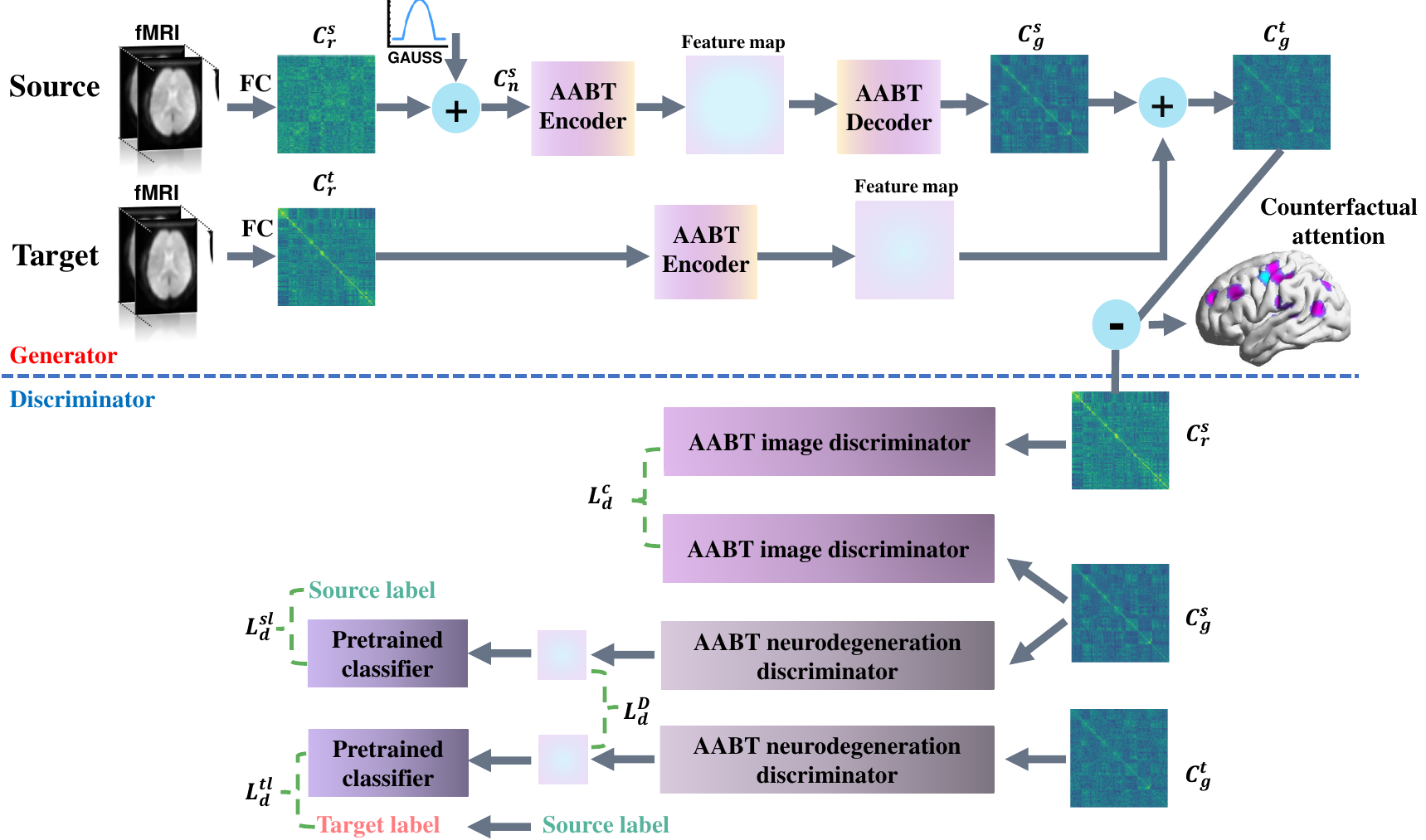}
    \vspace{-3mm}
    \caption{
    The proposed GCAN is illustrated as follows: The generator starts with a noisy source label input, $C^{s}_{n}$, using it to reconstruct the source label's FC. 
    It also extracts disease-related causal information from the dataset's mean target label FC, $C^{t}_{r}$. This causal information is then leveraged to transform the reconstructed source label FC into the target label FC, $C^{t}_{g}$. On the other side, the discriminator employs components sensitive to both imaging features and neurodegenerative indicators to verify that $C^{t}_{g}$ not only mimics the FC characteristics accurately but also encapsulates distinct cognitive information differentiating it from $C^{s}_{r}$.
    }
    \label{fig: Fig2}
    \vspace{-3mm}
\end{figure*}

\vspace{-4mm}
\subsubsection{Generator} 

As illustrated in Fig. \ref{fig: Fig2}, the generator begins by combining Gaussian noise and the FC matrix with the source label, denoted as input $C_n^{s}$. This input is then transformed into a feature map by the AABT encoder. Following this transformation, an AABT decoder, which mirrors the AABT encoder in architecture, reconstructs the feature map into a generated FC matrix, represented as $C_g^{s}$.
To derive counterfactual attention, the mean FC matrix associated with the target label undergoes encoding by the AABT to yield an average feature map of the target class, denoted as $C_r^{t}$. This map contains the causal information necessary for counterfactual inference. It is then combined with $C_g^{s}$ to facilitate the inference of the target FC matrix, denoted as $C_g^{t}$.
To ensure $C_g^{s}$ and $C_g^{t}$ closely mirror FC matrices, the generator loss includes perceptual ($L_p$), generative ($L_g$), and label cross-entropy ($L_c$) losses.
$L_p$ and $L_g$ maintain FC characteristics for $C_g^{s}$, while $L_c$ ensures $C_g^{t}$'s accuracy in target labeling, verified by a pre-trained classifier ($CLS$).
The total loss, $L_G$, is defined as follows:
    \begin{equation}
       L_p = MSE\left(VGG\left(C_g^{s}\right),VGG\left(C_r^{s}\right)\right),
    \label{eq:l_p}
    \end{equation}
    \vspace{-2mm}
    \begin{equation}
       L_{gen} = MSE\left(C_g^{s},C_r^{s}\right),
       \label{eq:l_g}
    \end{equation}
     \vspace{-2mm}
    \begin{equation}
       L_c = CE\left(CLS\left(C_g^{t}\right),y_t\right),
       \label{eq:l_c}
    \end{equation}
    \vspace{-2mm}
        \begin{equation}
       L_G = L_p + L_c + L_{gen}, 
       \label{eq:L_g}
    \end{equation}
where $MSE$ represents the mean square error, $VGG$ represents VGG16 network, and $CE$ represents cross-entropy, $y_t$ represents the expected target label.

\vspace{-5mm}
\subsubsection{Discriminator} 

The discriminator of GCAN, detailed in Fig. \ref{fig: Fig2}, ensures the fidelity of $C_g^{t}$ and $C_g^{s}$ through two main components: AABT image and AABT neurodegeneration. The image component evaluates $C_g^{s}$'s FC characteristics align with $C_r^{s}$, indicated by loss $L_d^{c}$. The neurodegeneration component focuses on identifying cognitive decline features within FC matrices, aiding in accurate subject classification by a pre-trained classifier. It employs cross-entropy loss $L_{d}^{sl}$ to align features of $C_g^{s}$ with the source label and $L_{d}^{tl}$ for matching $C_g^{t}$ features with the target label. Additionally, mean square error $L_{d}^{D}$ measures the distinction between feature maps of $C_g^{s}$ and $C_g^{t}$, promoting discriminative learning of cognitive features.

\vspace{-3mm}
    \begin{equation}
       L_d^{c} = mean\left\{log\left[1-S\left(D_i\left(C_g^{s}\right)\right)\right]*log\left[S\left(D_i\left(C_r^{s}\right)\right)\right]\right\}
    \label{eq:l_dc}
    \end{equation}
    \vspace{-3mm}
    \begin{equation}
       L_{d}^{sl} = CE\left(CLS\left(D_n\left(C_g^{s}\right)\right),y_s\right)
       \label{eq:l_dsl}
    \end{equation}
    \vspace{-3mm}
    \begin{equation}
       L_{d}^{tl} = CE\left(CLS\left(D_n\left(C_g^{t}\right)\right),l_t\right)
       \label{eq:l_dtl}
    \end{equation}
    \vspace{-3mm}
    \begin{equation}
       L_{d}^{D} = mean\left\{log\left[1-S\left(D_i\left(C_g^{s}\right)\right)\right]*log\left[1-S\left(D_i\left(C_g^{t}\right)\right)\right]\right\} 
       \label{eq:l_d}
    \end{equation}
    \vspace{-2mm}
    \begin{equation}
       L_D = L_d^{c} + L_{d}^{sl} + L_{d}^{tl} + L_{d}^{D}
       \label{eq:L_D}
    \vspace{-1mm}
    \end{equation}
where $D_i$ represents the image discriminator and $D_n$ represents the neurodegeneration discriminator, $y_s$ represents the source label.

\vspace{-2mm}
\subsection{Atlas-aware Bidirectional Transformer}
        \vspace{-2mm}

\begin{figure*}[h]
        \vspace{-3mm}
    \centering
    \includegraphics[scale=.38]{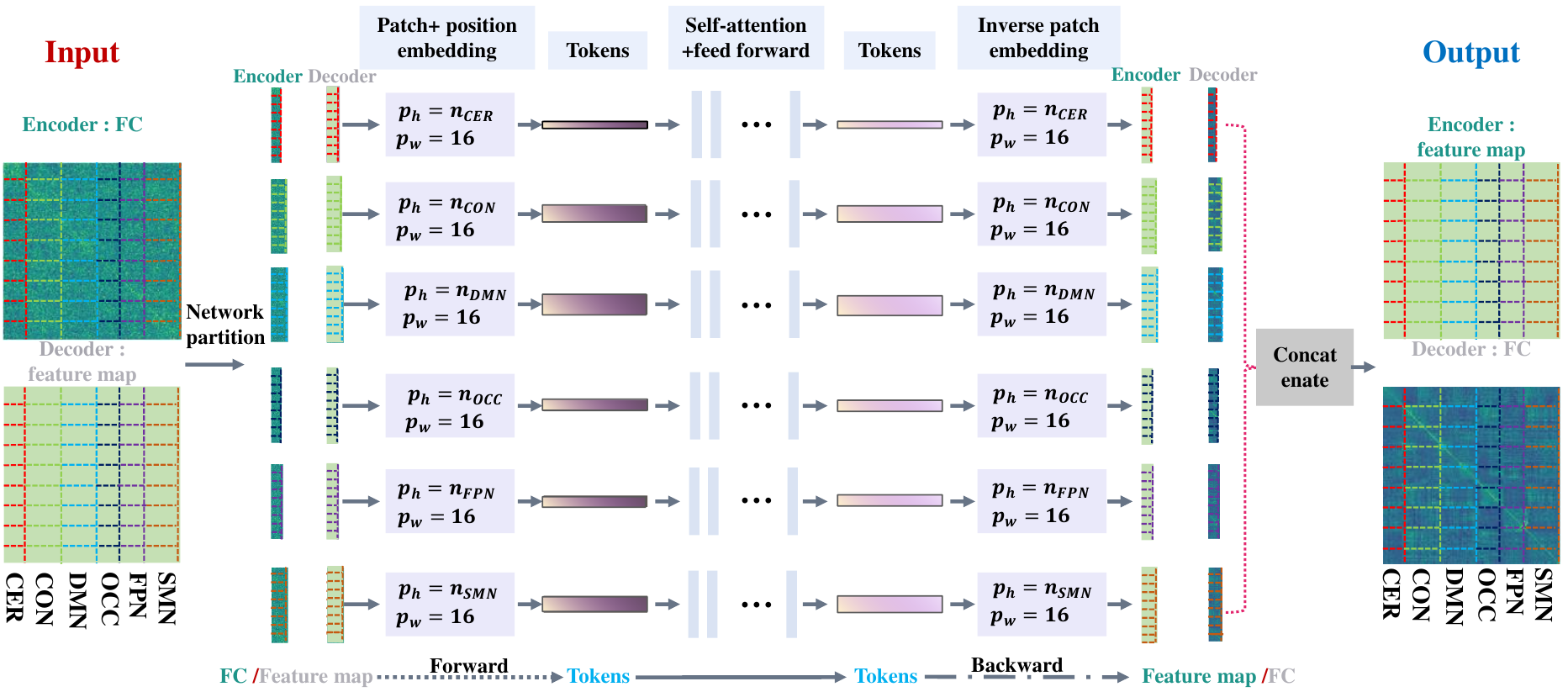}
        \vspace{-3mm}
    \caption{
    The \textbf{\textit{AABT}} mechanism involves the dynamic configuration of patch embedding and inverse patch embedding, contingent on the network's region count, facilitating both forward and backward processes. Here, $p_w$ and $p_h$ specify patch width and height, respectively, while $n_*$ indicates the network's region total.
    }
    \label{fig: Fig3}
    \vspace{-6mm}
\end{figure*}


\vspace{0mm}
Directly inputting the FC of all atlas networks into the Transformer block could result in uniform attention across all region correlations. Inspired by the vision Transformer \cite{Dosovitskiy2021vit} and the atlas-aware framework in fMRI \cite{Bannadabhavi2023com}, the basic block of AABT is designed to dynamically focus on the correlation of the current atlas network. The bidirectional structure including patch embedding and inverse patch embedding is employed to encode and decode the correlation in the Transformer. The patch embedding serves as the forward process to encode the input (FC or feature map) into tokens, while inverse patch embedding acts as the backward process to decode the tokens into outputs (FC or feature map). The combination of patch embedding and inverse patch embedding aims at encoding and decoding the current network correlation with a global perspective, aiding the model in understanding the current network correlation during both encoding and decoding processes. 
Initially, the input FC is divided into networks, including CER (cerebellum network), CON (cingulo-opercular network), DMN (default mode network), OCC (occipital network), FPN (fronto-parietal network), and SMN (sensorimotor network). 
During patch and position embedding, patch height matches each network's region count, with patch width set at 16. Tokens from these embeddings undergo self-attention and feed-forward blocks for encoding.
For decoding, inverse patch embedding recreates the FC or feature map.
The FC decoding relies on a narrowly focused convolutional network \cite{Tan2021gen}.
Inverse patch embedding decodes FC in the Transformer, offering a global view, as shown in Fig. \ref{fig: Fig3}. The output $O$ of AABT is defined as follows:
\vspace{-2mm}
\begin{equation}
       O = C\left\{\tilde{P}_i \left[SA_i\left(P_i\left(S_i\left(I\right)\right)\right)\right]\right\}  
    \label{eq:l_AABT}
    \vspace{-2mm} 
    \end{equation}
where $I$ represents the input feature map or FC, $i$ represents $i$-th network of FC, $S_i$ represents network segmentation, $P_i$ represents the patch embedding and position embedding operation, $SA$ represents self attention and feed forward, $\tilde{P}_i$ represents inverse patch embedding, $C$ represents concatenation operation.

\vspace{0mm}
\section{Experiments and Results}
\vspace{0mm}
\subsection{Dataset and Experiment Setup}
\vspace{0mm}
\subsubsection{Dataset} 
In this study, both hospital-collected data and Alzheimer's Disease Neuroimaging Initiative (ADNI) data are employed to train and validate the proposed method. The hospital-collected data comprises 77 HC, 75 SCD patients, and 99 MCI patients. 
The ADNI data consists of 67 HC, 22 SCD patients, and 95 MCI patients. The data undergo preprocessing using SPM12 \cite{Tzourio-Mazoyer2002spm}, including slice-timing correction, head motion estimation and correction, intra-subject registration, and co-registration.

\vspace{-3mm}
\subsubsection{Experiment Setup} The depth of the Transformer in the encoder and decoder of the generator is set to 3, while in the image and neurodegeneration discriminator part, it is set to 8. Other hyperparameters of the model can be found in the code. 
The performance of the pre-trained and final classifiers is evaluated using accuracy (ACC), recall, precision, and F1-score (F1). 

\vspace{-2mm}
\subsection{Results}
\vspace{-2mm}

\subsubsection{Diagnostic Performance} 
To validate the diagnostic performance, baseline models are constructed based on the following formulations:
\begin{itemize}
\vspace{-1mm}
\item[$\bullet$] 
$R*$ represents either ResNet10 or ResNet18 \cite{Szegedy2016Res}. $T*$ symbolizes a Transformer with varied multi-head self-attention counts. 'B' means 16 Transformer heads, 'L' for 32, and 'S' for 8. 'A' marks the addition of channel attention, a prevalent attention mechanism \cite{Woo2018ca}.
\item[$\bullet$] $R*//$ indicates diagnostic model solely constructed by ResNet. $RA$ signifies the diagnostic model constructed using ResNet and channel attention. $RT*$ denotes the diagnostic model constructed using ResNet and Transformer.
\end{itemize}
\vspace{-1mm}
The proposed method employs ResNet10 and Transformer with 16 heads. While the baseline model directly inputs FC, the proposed method inputs FC masked by counterfactual attention. As demonstrated in Table \ref{tab: performance}, the proposed method achieves superior diagnostic performance across three tasks and two datasets.

\begin{table}[h]
\vspace{-1mm}
\caption{The diagnostic performance of binary diagnostic task (HC vs. SCD, HC vs. MCI, and SCD vs. MCI) on hospital-collected and ADNI datasets.}
\vspace{-4mm}
\resizebox{\linewidth}{!}{
\begin{tabular}{ccc|cccccccc}
\hline
\multicolumn{1}{l}{} &
  \multicolumn{1}{l}{} &
  \multicolumn{1}{l|}{} &
  \multicolumn{4}{c|}{Hospital} &
  \multicolumn{4}{c}{ADNI} \\ \cline{4-11} 
\multicolumn{1}{l}{} &
  \multicolumn{1}{l}{} &
   &
  Acc &
  Recall &
  Precision &
  \multicolumn{1}{c|}{F1} &
  Acc &
  Recall &
  Precision &
  F1 \\ \hline
\multicolumn{1}{c|}{\multirow{11}{*}{\begin{tabular}[c]{@{}c@{}}HC\\ vs.\\ SCD\end{tabular}}} &
  \multicolumn{1}{c|}{\multirow{5}{*}{R10}} &
  // &
  0.8333 &
  0.4667 &
  1.0000 &
  \multicolumn{1}{c|}{0.6364} &
  0.6786 &
  \textbf{0.7823} &
  0.6130 &
  0.6874 \\
\multicolumn{1}{c|}{} &
  \multicolumn{1}{c|}{} &
  A &
  0.8667 &
  0.6667 &
  0.8000 &
  \multicolumn{1}{c|}{0.7273} &
  0.6905 &
  0.8413 &
  0.6033 &
  0.7027 \\
\multicolumn{1}{c|}{} &
  \multicolumn{1}{c|}{} &
  T-S &
  0.8000 &
  0.4000 &
  0.8000 &
  \multicolumn{1}{c|}{0.5333} &
  0.6190 &
  0.3764 &
  0.6667 &
  0.4812 \\
\multicolumn{1}{c|}{} &
  \multicolumn{1}{c|}{} &
  T-B &
  0.8000 &
  0.4000 &
  0.8000 &
  \multicolumn{1}{c|}{0.5333} &
  0.6667 &
  0.6049 &
  0.6627 &
  0.6325 \\
\multicolumn{1}{c|}{} &
  \multicolumn{1}{c|}{} &
  T-L &
  0.8000 &
  0.4000 &
  0.8000 &
  \multicolumn{1}{c|}{0.5333} &
  0.6049 &
  0.7078 &
  0.5802 &
  0.6377 \\ \cline{2-11} 
\multicolumn{1}{c|}{} &
  \multicolumn{1}{c|}{\multirow{5}{*}{R18}} &
  // &
  0.8000 &
  0.4000 &
  0.8000 &
  \multicolumn{1}{c|}{0.5333} &
  0.6905 &
  0.5510 &
  0.6648 &
  0.6026 \\
\multicolumn{1}{c|}{} &
  \multicolumn{1}{c|}{} &
  A &
  0.8000 &
  0.4000 &
  0.8000 &
  \multicolumn{1}{c|}{0.5333} &
  0.6310 &
  0.3220 &
  0.7143 &
  0.4439 \\
\multicolumn{1}{c|}{} &
  \multicolumn{1}{c|}{} &
  T-S &
  0.8000 &
  0.4000 &
  0.8000 &
  \multicolumn{1}{c|}{0.5333} &
  0.6207 &
  0.3739 &
  0.6345 &
  0.4705 \\
\multicolumn{1}{c|}{} &
  \multicolumn{1}{c|}{} &
  T-B &
  0.8000 &
  0.5333 &
  0.6800 &
  \multicolumn{1}{c|}{0.5978} &
  0.6543 &
  0.3128 &
  0.6614 &
  0.4247 \\
\multicolumn{1}{c|}{} &
  \multicolumn{1}{c|}{} &
  T-L &
  0.8000 &
  0.4000 &
  0.8000 &
  \multicolumn{1}{c|}{0.5333} &
  0.6420 &
  0.4815 &
  0.6587 &
  0.5563 \\ \cline{2-11} 
\multicolumn{1}{c|}{} &
  \multicolumn{1}{c|}{Proposed} &
   &
  \textbf{0.9333} &
  \textbf{0.8667} &
  \textbf{1.0000} &
  \multicolumn{1}{c|}{\textbf{0.9286}} &
  \textbf{0.7284} &
  0.6667 &
  \textbf{0.7445} &
  \textbf{0.7035} \\ \hline
\multicolumn{1}{c|}{\multirow{11}{*}{\begin{tabular}[c]{@{}c@{}}HC\\ vs.\\ MCI\end{tabular}}} &
  \multicolumn{1}{c|}{\multirow{5}{*}{R10}} &
  // &
  0.6552 &
  0.9004 &
  0.6552 &
  \multicolumn{1}{c|}{0.7585} &
  0.6562 &
  0.8611 &
  0.6456 &
  0.7379 \\
\multicolumn{1}{c|}{} &
  \multicolumn{1}{c|}{} &
  A &
  0.6458 &
  0.9167 &
  0.6331 &
  \multicolumn{1}{c|}{0.7490} &
  0.6207 &
  \textbf{0.9632} &
  0.6137 &
  0.7497 \\
\multicolumn{1}{c|}{} &
  \multicolumn{1}{c|}{} &
  T-S &
  0.6437 &
  0.8812 &
  0.6381 &
  \multicolumn{1}{c|}{0.7402} &
  0.6146 &
  0.8542 &
  0.6102 &
  0.7119 \\
\multicolumn{1}{c|}{} &
  \multicolumn{1}{c|}{} &
  T-B &
  0.6322 &
  0.8084 &
  0.6528 &
  \multicolumn{1}{c|}{0.7223} &
  0.6458 &
  0.9514 &
  0.6193 &
  0.7502 \\
\multicolumn{1}{c|}{} &
  \multicolumn{1}{c|}{} &
  T-L &
  0.6667 &
  0.9195 &
  0.6468 &
  \multicolumn{1}{c|}{0.7594} &
  0.6465 &
  0.8364 &
  0.6505 &
  0.7318 \\ \cline{2-11} 
\multicolumn{1}{c|}{} &
  \multicolumn{1}{c|}{\multirow{5}{*}{R18}} &
  // &
  0.7126 &
  0.9080 &
  0.7125 &
  \multicolumn{1}{c|}{0.7985} &
  0.6667 &
  0.6929 &
  0.6729 &
  0.6828 \\
\multicolumn{1}{c|}{} &
  \multicolumn{1}{c|}{} &
  A &
  0.7011 &
  0.7739 &
  \textbf{0.7755} &
  \multicolumn{1}{c|}{0.7747} &
  0.6458 &
  0.7986 &
  0.6468 &
  0.7147 \\
\multicolumn{1}{c|}{} &
  \multicolumn{1}{c|}{} &
  T-S &
  0.6782 &
  0.8889 &
  0.6629 &
  \multicolumn{1}{c|}{0.7594} &
  0.6458 &
  0.8194 &
  0.6610 &
  0.7317 \\
\multicolumn{1}{c|}{} &
  \multicolumn{1}{c|}{} &
  T-B &
  0.6437 &
  \textbf{1.0000} &
  0.6247 &
  \multicolumn{1}{c|}{0.7690} &
  0.6667 &
  0.7847 &
  0.6724 &
  0.7242 \\
\multicolumn{1}{c|}{} &
  \multicolumn{1}{c|}{} &
  T-L &
  0.6207 &
  0.8927 &
  0.6243 &
  \multicolumn{1}{c|}{0.7348} &
  0.6354 &
  0.6736 &
  \textbf{0.7037} &
  0.6883 \\ \cline{2-11} 
\multicolumn{1}{c|}{} &
  \multicolumn{1}{c|}{Proposed} &
   &
  \textbf{0.7471} &
   0.9816 &
  0.7056 &
  \multicolumn{1}{c|}{\textbf{0.8210}} &
  \textbf{0.6970} &
  0.8653 &
  0.6709 &
  \textbf{0.7558} \\ \hline
\multicolumn{1}{c|}{\multirow{11}{*}{\begin{tabular}[c]{@{}c@{}}SCD\\ vs. \\MCI\end{tabular}}} &
  \multicolumn{1}{c|}{\multirow{5}{*}{R10}} &
  // &
  0.7778 &
  0.9392 &
  0.8133 &
  \multicolumn{1}{c|}{0.8717} &
  0.6989 &
  0.7233 &
  \textbf{0.7610} &
  0.7417 \\
\multicolumn{1}{c|}{} &
  \multicolumn{1}{c|}{} &
  A &
  0.8500 &
  \textbf{1.0000} &
  0.8500 &
  \multicolumn{1}{c|}{0.9189} &
  0.6989 &
  0.7247 &
  0.7552 &
  0.7396 \\
\multicolumn{1}{c|}{} &
  \multicolumn{1}{c|}{} &
  T-S &
  0.7692 &
  1.0000 &
  0.7692 &
  \multicolumn{1}{c|}{0.8695} &
  0.7204 &
  0.7864 &
  0.7935 &
  0.7899 \\
\multicolumn{1}{c|}{} &
  \multicolumn{1}{c|}{} &
  T-B &
  0.7692 &
  1.0000 &
  0.7692 &
  \multicolumn{1}{c|}{0.8695} &
  0.6989 &
  0.9541 &
  0.6726 &
  0.7890 \\
\multicolumn{1}{c|}{} &
  \multicolumn{1}{c|}{} &
  T-L &
  0.7949 &
  1.0000 &
  0.7857 &
  \multicolumn{1}{c|}{0.8800} &
  0.6989 &
  0.8380 &
  0.7097 &
  0.7685 \\ \cline{2-11} 
\multicolumn{1}{c|}{} &
  \multicolumn{1}{c|}{\multirow{5}{*}{R18}} &
  // &
  0.7949 &
  1.0000 &
  0.7857 &
  \multicolumn{1}{c|}{0.8800} &
  0.6774 &
  \textbf{0.9828} &
  0.6523 &
  0.7841 \\
\multicolumn{1}{c|}{} &
  \multicolumn{1}{c|}{} &
  A &
  0.8500 &
  1.0000 &
  0.8500 &
  \multicolumn{1}{c|}{0.9189} &
  0.7097 &
  0.8165 &
  0.7322 &
  0.7721 \\
\multicolumn{1}{c|}{} &
  \multicolumn{1}{c|}{} &
  T-S &
  0.7692 &
  1.0000 &
  0.7692 &
  \multicolumn{1}{c|}{0.8695} &
  0.6989 &
  0.8509 &
  0.6943 &
  0.7647 \\
\multicolumn{1}{c|}{} &
  \multicolumn{1}{c|}{} &
  T-B &
  0.8205 &
  1.0000 &
  0.8047 &
  \multicolumn{1}{c|}{0.8918} &
  0.6667 &
  0.6344 &
  0.7586 &
  0.6910 \\
\multicolumn{1}{c|}{} &
  \multicolumn{1}{c|}{} &
  T-L &
  0.7692 &
  1.0000 &
  0.7692 &
  \multicolumn{1}{c|}{0.8695} &
  0.6882 &
  0.9140 &
  0.6779 &
  0.7784 \\ \cline{2-11} 
\multicolumn{1}{c|}{} &
  \multicolumn{1}{l|}{Proposed} &
   &
  \textbf{0.9487} &
  0.9707 &
  \textbf{0.9615} &
  \multicolumn{1}{c|}{\textbf{0.9661}} &
  \textbf{0.7312} &
  0.8656 &
  0.7307 &
  \textbf{0.7924} \\ \hline
\end{tabular}
}
\label{tab: performance}
\vspace{-5mm}
\end{table}
\vspace{-0mm}

\begin{table}[h]
    \caption{The network of top 10 regions at the counterfactual attention in HC vs. MCI, HC vs. SCD, and SCD vs. MCI diagnostic task.
    }
    \vspace{-2mm}
\resizebox{\linewidth}{!}{
\begin{tabular}{c|c|cccccccccc}
\hline
\multirow{2}{*}{HC vs. SCD}  &  $+$  & FPN & DMN & SMN & CON & OCC & SMN & OCC & CON & CER & FPN \\ \cline{2-12} 
                         & $-$& FPN & DMN & SMN & OCC & CON & SMN & FPN & CON & DMN & SMN \\ \hline
\multirow{2}{*}{HC vs. MCI}  & $+$& FPN & SMN & FPN & OCC & SMN & DMN & CER & OCC & DMN & FPN \\ \cline{2-12} 
                         & $-$& FPN & SMN & FPN & OCC & SMN & FPN & DMN & CON & CER & OCC \\ \hline
\multirow{2}{*}{SCD vs. MCI} & $+$& DMN & SMN & CON & DMN & FPN & DMN & FPN & CER & SMN & DMN \\ \cline{2-12} 
                         &  $-$ & FPN & DMN & CER & SMN & SMN & SMN & CON & DMN & CON & OCC \\ \hline
\end{tabular}
}
\label{tab: net}

    \vspace{-0mm}
\end{table}

\begin{figure*}[h]  
    \centering
    \includegraphics[scale=.47]{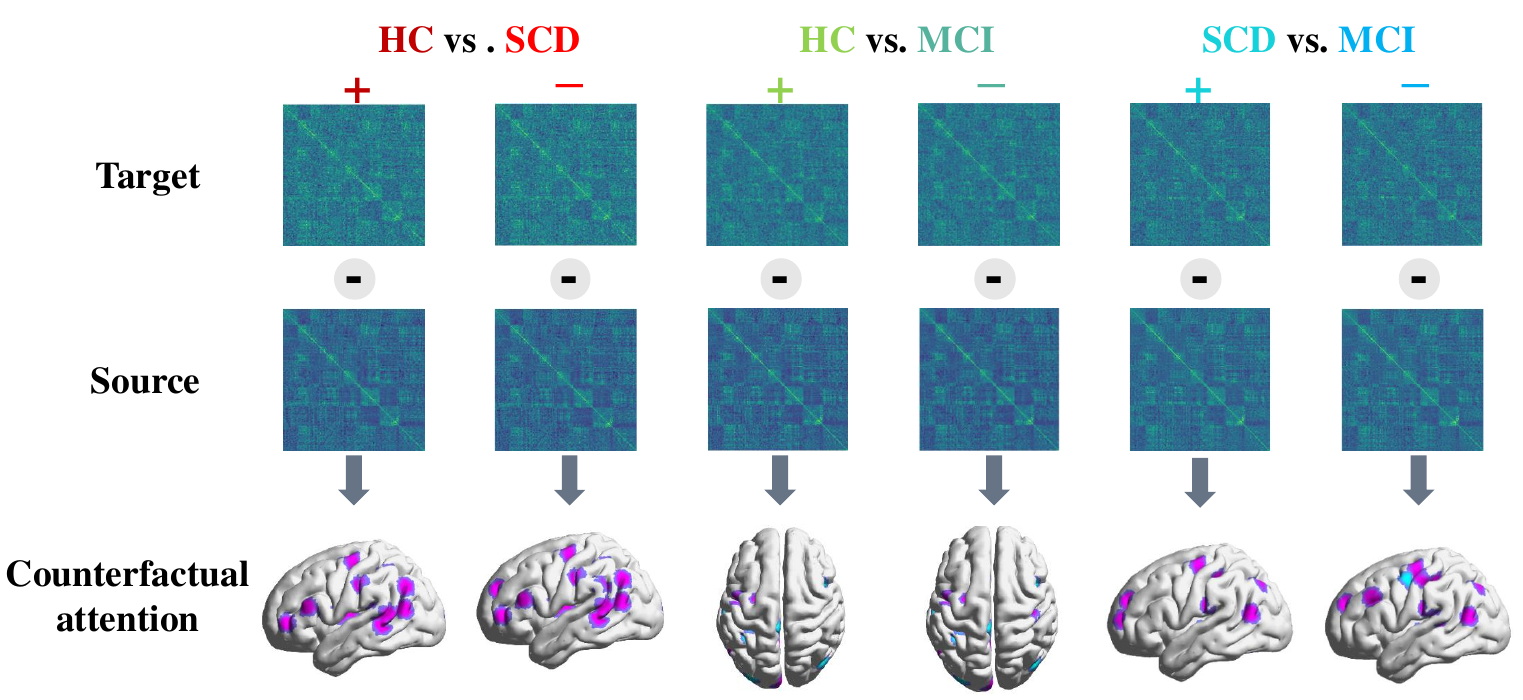}
    \vspace{-5mm}
    \caption{The generated target label FC and counterfactual attention in HC vs. MCI, HC vs. SCD, and SCD vs. MCI diagnostic task.
    }
    \label{fig: Fig5}
\end{figure*}
\vspace{000mm}

\vspace{-3mm}
\subsubsection{Counterfactual Attention Map}
During the neurodegeneration process from HC to MCI, significant FC changes occur in regions such as the prefrontal cortex, cingulate cortex, and hippocampus \cite{Jin2020hcmci}. Similarly, during intervention in MCI, significant changes are observed in the FC of regions like the cingulate cortex and gyrus \cite{Eyre2016intervention}. Hence, throughout the conversion process of each binary diagnostic task, most attention regions should remain consistent with slight variations.
In Fig. \ref{fig: Fig5}, the attention of each FC region is calculated and depicted using BrainNet Viewer \cite{Xia2023brain}. Prominent attention regions across different conversion processes in diagnostic tasks, including HC vs. MCI, HC vs. SCD, and SCD vs. MCI, show considerable overlap, although some regional differences exist. These findings align closely with previously reported research. The networks of the top 10 regions in the counterfactual attention are outlined in Table \ref{tab: net}. These networks are highly associated with cognitive decline \cite{Mah2021network,Ghanbari2023network}. Hence, the significant regions identified by counterfactual attention are strongly linked to neurodegeneration.

\vspace{-5mm}
\subsubsection{Ablation study} To validate the benefits of counterfactual attention, we conduct an ablation study on the same diagnostic model. One model inputs FC directly, while the other inputs FC masked by counterfactual attention. As depicted in Table \ref{tab: ab}, the model utilizing counterfactual attention has superior diagnostic performance across three tasks and two datasets.

\begin{table}[H]
\vspace{-1mm}

\caption{Ablation study on hospital-collected and ADNI datasets.
    }
    \vspace{-2mm}
\resizebox{\linewidth}{!}{
\begin{tabular}{c|c|cccc|cccc}
\hline
\multirow{2}{*}{} & \multirow{2}{*}{\begin{tabular}[c]{@{}c@{}}Counterfactual \\ attention\end{tabular}} & \multicolumn{4}{c|}{Hospital} & \multicolumn{4}{c}{ADNI} \\ \cline{3-10} 
                          &  & Acc     & Recall & Precision & F1     & Acc    & Recall & Precision & F1     \\ \hline
\multirow{2}{*}{HC vs. SCD}  &\ding{55}  & 0.80000 & 0.4000 & 0.8000    & 0.5333 & 0.6667 & 0.6049 & 0.6627    & 0.6325 \\
                          & \ding{52}  & \textbf{0.9333}  & \textbf{0.8667} & \textbf{1.0000}    & \textbf{0.9286} & \textbf{0.7284} & \textbf{0.6667} & \textbf{0.7445}    & \textbf{0.7035} \\ \hline
\multirow{2}{*}{HC vs. MCI}  & \ding{55}  & 0.6322  & 0.8084 & 0.6528    & 0.7223 & 0.6458 & 0.9514 & 0.6193    & 0.7502 \\
                          & \ding{52} & \textbf{0.7471}  & \textbf{0.9816} & \textbf{0.7056}    & \textbf{0.8210}  & \textbf{0.6970}  & \textbf{0.8653} & \textbf{0.6709}    & \textbf{0.7558} \\ \hline
\multirow{2}{*}{MCI vs. SCD} & \ding{55}  & 0.7692  & \textbf{1.0000} & 0.7692    & 0.8695 & 0.6989 & \textbf{0.9541} & 0.6726    & 0.7890  \\
                          & \ding{52} & \textbf{0.9487}  & 0.9707 & \textbf{0.9615}    & \textbf{0.9661} & \textbf{0.7312} & 0.8656 & \textbf{0.7307}    & \textbf{0.7924} \\ \hline
\end{tabular}
}
 \vspace{-5mm}
\label{tab: ab}
\end{table}

\vspace{-8mm}
\section{Conclusion}
\vspace{-3mm}
To improve both the explainability and performance of the FC-based SCD/MCI diagnostic model, we propose the GCAN, which directs the model's attention towards regions associated with neurodegeneration, termed the 'counterfactual attention map'. 
This objective is achieved by constructing the generator and the discriminator with AABT, enabling the generation of the target label FC and the subtraction of the source label FC. AABT adapts an atlas-aware bidirectional transformer and offers global insights into the target label FC reconstruction.
Experimental results confirm that the counterfactual attention map aligns with empirical observations and domain knowledge of SCD and MCI, which demonstrates the explainability of the proposed GCAN. The diagnostic performance and ablation study demonstrate the effectiveness of counterfactual attention. 

\subsubsection{Acknowledgement} 
This work was founded by the National Natural Science Foundation of China (Grants 32361143787), the China Postdoctoral Science Foundation (Grants 2023M730873, GZB20230960). We have to appreciate the hispital-collected dataset provided by Hospital of Guangxi University of Traditional Chinese Medicine.


\begin{thebibliography}{8}

\bibitem{Ramírez-Toraño2022}
Ramírez-Toraño, F., Bruña, R., Bruña, R., Bruña, R., Frutos-Lucas, J.D., Frutos-Lucas, J.D., Rodríguez-Rojo, I.C., Pedro, S.M., Pedro, S.M., Delgado-Losada, M.L., Gómez-Ruiz, N., Barabash, A., Marcos, A., Higes, R., Maestú, F., Maestú, F., Maestú, F.: Functional Connectivity Hypersynchronization in Relatives of Alzheimer's Disease Patients: An Early E/I Balance Dysfunction?. Cereb. Cortex \textbf{31}(2), 1201--1210 (2020) 

\bibitem{Liebe2022FC}
Liebe, T., Dordevic, M., Kaufmann, J., Avetisyan, A., Skalej, M., Müller, N.G.: Investigation of The Functional Pathogenesis of Mild Cognitive Impairment by Localisation-based Locus Coeruleus Resting-state fMRI. Hum. Brain. Mapp \textbf{43}(18), 5630--5642 (2022)

\bibitem{Li2021CNN}
Li, Y., Liu, J., Jiang, Y., Liu, Y., Lei, B.: Virtual Adversarial Training-Based Deep Feature Aggregation Network From Dynamic Effective Connectivity for MCI Identification. IEEE Trans. Med. Imaging \textbf{41}(1), 237--251 (2021)


\bibitem{Zuo2023t}
Zuo, Q., Zhong, N., Pan, Y., Wu, H., Lei, B., Wang, S.: Brain Structure-Function Fusing Representation Learning Using Adversarial Decomposed-VAE for Analyzing MCI. IEEE Trans. Neural Syst. Rehabilitation Eng. \textbf{31}, 4017--4028 (2023)


\bibitem{Selvaraju2016gradcam}
Selvaraju, R.R., Das, A., Vedantam, R., Cogswell, M., Parikh, D., Batra, D.: Grad-CAM: Visual Explanations from Deep Networks via Gradient-Based Localization. Int J Comput Vision \textbf{128}(2), 336--359 (2016)

\bibitem{Wang2020scorecam}
Wang, H., Wang, Z., Du, M., Yang, F., Zhang, Z., Ding, S., Mardziel, P., Hu, X.: Score-CAM: Score-Weighted Visual Explanations for Convolutional Neural Networks. In 2020 IEEE/CVF Conference on Computer Vision and Pattern Recognition Workshops (CVPRW), pp. 111--119. IEEE, Seattle, The United States of America, (2019). \doi{10.1109/CVPRW50498.2020.00020}

\bibitem{Oh2021cm}
Oh, K., Yoon, J., Suk, H.: Learn-Explain-Reinforce: Counterfactual Reasoning and its Guidance to Reinforce an Alzheimer's Disease Diagnosis Model. IEEE Trans. Pattern. Anal. \textbf{45}(4), 4843--4857 (2021)

\bibitem{Ren2023cm}
Ren, Z., Sun, Y., Wang, M., Feng, Y., Li, X., Jin, C., Yang, J., Lian, C., Wang, F.: Punctate White Matter Lesion Segmentation in Preterm Infants Powered by Counterfactually Generative Learning. In: Medical Image Computing and Computer Assisted Intervention–MICCAI 2023: 26th International Conference, pp. 220--229. Springer, Vancouver, Canada (2023). \doi{10.1007/978-3-031-43904-9_22}

\bibitem{Dosovitskiy2021vit}
Dosovitskiy, A., Beyer, L., Kolesnikov, A., Weissenborn, D., Zhai, X., Unterthiner, T., Dehghani, M., Minderer, M., Heigold, G., Gelly, S., Uszkoreit, J., Houlsby, N.: An Image is Worth 16x16 Words: Transformers for Image Recognition at Scale. ArXiv, abs/2010.11929, (2020)

\bibitem{Bannadabhavi2023com}
Bannadabhavi, A., Lee, S., Deng, W., Li, X.: Community-Aware Transformer for Autism Prediction in fMRI Connectome. In: Medical Image Computing and Computer Assisted Intervention–MICCAI 2023: 26th International Conference, vol. 14227, pp. 287--297. Springer, Vancouver, Canada (2023). \doi{10.1007/978-3-031-43993-3_28}

\bibitem{Tan2021gen}
Tan, Y., Ting, C., Noman, F.M., Phan, R.C., Ombao, H.C.: A Unified Framework for Static and Dynamic Functional Connectivity Augmentation for Multi-Domain Brain Disorder Classification. In: 2023 IEEE International Conference on Image Processing (ICIP), pp. 635--639. IEEE, Kuala Lumpur, Malaysia (2023). \doi{10.1109/ICIP49359.2023.10222266}

\bibitem{Tzourio-Mazoyer2002spm}
Tzourio-Mazoyer, N., Landeau, B., Papathanassiou, D., Crivello, F., Etard, O., Delcroix, N., Mazoyer, B., Joliot, M.: Automated Anatomical Labeling of Activations in SPM Using a Macroscopic Anatomical Parcellation of the MNI MRI Single-Subject Brain. Neuroimage \textbf{15}(1), 273--289 (2002)

\bibitem{Szegedy2016Res}
Szegedy, C., Ioffe, S., Vanhoucke, V., Alemi, A.A.: Inception-v4, Inception-ResNet and the Impact of Residual Connections on Learning. In: 31st AAAI Conference on Artificial Intelligence, pp. 4278--4284. Association Advancement Artificial Intelligence, San Francisco, The United States of America (2016). \doi{10.1609/aaai.v31i1.11231}

\bibitem{Woo2018ca}
Woo, S., Park, J., Lee, J., Kweon, I.: CBAM: Convolutional Block Attention Module. In: 15th European Conference on Computer Vision (ECCV), pp. 3-19. Sringer, Munich, Germany (2018). \doi{10.1007/978-3-030-01234-2_1}

\bibitem{Jin2020hcmci}
Jin, D., Wang, P., Zalesky, A., Liu, B., Song, C., Wang, D., Xu, K., Yang, H., Zhang, Z., Yao, H., Zhou, B., Han, T., Zuo, N., Han, Y., Lu, J., Wang, Q., Yu, C., Zhang, X., Zhang, X., Jiang, T., Zhou, Y., Liu, Y.: Grab‐AD: Generalizability and Reproducibility of Altered Brain Activity and Diagnostic Classification in Alzheimer's Disease. Hum. Brain. Mapp. \textbf{41}(12), 3379--3391 (2020)

\bibitem{Eyre2016intervention}
Eyre, H. A., Acevedo, B., Yang, H., Siddarth, P., Van Dyk, K., Ercoli, L., Leaver, A. M., Cyr, N. S., Narr, K., Baune, B. T., Khalsa, D. S., Lavretsky, H.: Changes in Neural Connectivity and Memory Following a Yoga Intervention for Older Adults: A Pilot Study. J. Alzheimer's Dis. \textbf{52}(2), 673–-684 (2016)

\bibitem{Xia2023brain}
Xia, M., Wang, J., He, Y.: BrainNet Viewer: A Network Visualization Tool for Human Brain Connectomics. PLoS One \textbf{8}7, e68910 (2013)

\bibitem{Mah2021network}
Mah, L., Murari, G., Vandermorris, S., Chen, J., Verhoeff, N.P., Herrmann, N.: Distinct Patterns of Posterior Default Mode Network-Medial Temporal Lobe Connectivity in Mild Cognitive Impairment and Subjective Cognitive Decline. Alzheimers. Dement. \textbf{17}(S4), e055832 (2021)

\bibitem{Ghanbari2023network}
Ghanbari, M., Li, G., Hsu, L., Yap, P.: Accumulation of Network Redundancy Marks the Early Stage of Alzheimer's Disease. Hum. Brain. Mapp. \textbf{44}(8), 2993--3006 (2023)





\end{thebibliography}
\end{document}